\documentstyle[12pt,aasms4]{article}
\begin{document}
\def\chaphead{}

\def\hut{Hubble type\ }
\def\vc{V$_{\rm C}$\ }
\def\mb{M$_{\rm B}$\ }
\def\av{A$_{\rm V}$\ }
\def\lamlam{$\lambda\lambda$}

\def\deg{$^\circ$}
\def\degrees{$^\circ$}
\def\Vlasov{collisionless Boltzmann\ }
\def\lsls{\ll}
\def\grgr{\gg}
\def\erf{\mathop{\rm erf}\nolimits} 
\def\eqv{\equiv}
\def\real{\Re e}
\def\imag{\Im m}
\def\ctrline#1{\centerline{#1}}
\def\spose#1{\hbox to 0pt{#1\hss}}
     
\def\={\overline}
\def\sections{\S}
\newcount\notenumber
\notenumber=1
\newcount\eqnumber
\eqnumber=1
\newcount\fignumber
\fignumber=1
\newbox\abstr
\newbox\figca     
\def\yyskip{\penalty-100\vskip6pt plus6pt minus4pt}
     
\def\numberpara{\yyskip\noindent}
     
\def\km{{\rm\,km}}
\def\kms{{\rm\ km\ s$^{-1}$}}
\def\kpc{{\rm\,kpc}}
\def\mpc{{\rm\,Mpc}}
\def\etal{{\it et al. }}
\def\eg{{\it e.g., }}
\def\ie{{\it i.e., }}
\def\cf{{\it cf. }}
\def\msun{{\rm\,M_\odot}}
\def\lsun{{\rm\,L_\odot}}
\def\rsun{{\rm\,R_\odot}}
\def\pc{{\rm\,pc}}
\def\cm{{\rm\,cm}}
\def\yr{{\rm\,yr}}
\def\au{{\rm\,AU}}
\def\AU{{\rm\,AU}}
\def\gm{{\rm\,g}}
\def\s{{\rmss}}
\def\dyne{{\rm\,dyne}}
     
\def\note#1{\footnote{$^{\the\notenumber}$}{#1}\global\advance\notenumber by 1}
\def\foot#1{\raise3pt\hbox{\eightrm \the\notenumber}
     \hfil\par\vskip3pt\hrule\vskip6pt
     \noindent\raise3pt\hbox{\eightrm \the\notenumber}
     #1\par\vskip6pt\hrule\vskip3pt\noindent\global\advance\notenumber by 1}
\def\propo{\propto}
\def\larrow{\leftarrow}
\def\rarrow{\rightarrow}
\def\sectionhead#1{\penalty-200\vskip24pt plus12pt minus6pt
        \centerline{\bbrm#1}\vskip6pt}
     
\def\Dt{\spose{\raise 1.5ex\hbox{\hskip3pt$\mathchar"201$}}}    
\def\dt{\spose{\raise 1.0ex\hbox{\hskip2pt$\mathchar"201$}}}    
\def\llangle{\langle\langle}
\def\rrangle{\rangle\rangle}
\def\ldotss{\ldots}
\def\del{\b\nabla}
     
\def\new{{\rm\chaphead\the\eqnumber}\global\advance\eqnumber by 1}
\def\ref#1{\advance\eqnumber by -#1 \chaphead\the\eqnumber
     \advance\eqnumber by #1 }
\def\last{\advance\eqnumber by -1 {\rm\chaphead\the\eqnumber}\advance
     \eqnumber by 1}
\def\eqnam#1{\xdef#1{\chaphead\the\eqnumber}}
     
\def\nfig{\chaphead\the\fignumber\global\advance\fignumber by 1}
\def\nfiga#1{\chaphead\the\fignumber{#1}\global\advance\fignumber by 1}
\def\rfig#1{\advance\fignumber by -#1 \chaphead\the\fignumber
     \advance\fignumber by #1}
\def\refindent{\par\noindent\parskip=4pt\hangindent=3pc\hangafter=1 }

\def\apj#1#2#3{\refindent#1,  {ApJ,\ }{#2}, #3}
\def\apjsup#1#2#3{\refindent#1,  {ApJS\ }{#2}, #3}
\def\aasup#1#2#3{\refindent#1,  { AA Sup.,\ }{#2}, #3}
\def\aas#1#2#3{\refindent#1,  { Bull. Am. Astr. Soc.,\ }{#2}, #3}
\def\apjlett#1#2#3{\refindent#1,  { ApJL,\  }{#2}, #3}
\def\mn#1#2#3{\refindent#1,  { MNRAS,\ }{#2}, #3}
\def\mnras#1#2#3{\refindent#1,  { M.N.R.A.S., }{#2}, #3}
\def\annrev#1#2#3{\refindent#1, { ARA \& A,\ }
{\bf2}, #3}
\def\aj#1#2#3{\refindent#1,  { AJ,\  }{#2}, #3}
\def\phrev#1#2#3{\refindent#1, { Phys. Rev.,}{#2}, #3}
\def\aa#1#2#3{\refindent#1,  { AA,\ }{#2}, #3}
\def\nature#1#2#3{\refindent#1,  { Nature,\ }{#2}, #3}
\def\icarus#1#2#3{\refindent#1,  { Icarus, }{#2}, #3}
\def\pasp#1#2#3{\refindent#1,  { PASP,\ }{#2}, #3}
\def\appopt#1#2#3{\refindent#1,  { App. Optics,\  }{#2}, #3}
\def\spie#1#2#3{\refindent#1,  { Proc. of SPIE,\  }{#2}, #3}
\def\opteng#1#2#3{\refindent#1,  { Opt. Eng.,\  }{#2}, #3}
\def\refpaper#1#2#3#4{\refindent#1,  { #2 }{#3}, #4}
\def\refbook#1{\refindent#1}
\def\science#1#2#3{\refindent#1, { Science, }{#2}, #3}
     
\def\chapbegin#1#2{\eject\vskip36pt\par\noindent{\chapheadfont#1\hskip30pt
     #2}\vskip36pt}
\def\sectionbegin#1{\vskip30pt\par\noindent{\bf#1}\par\vskip15pt}
\def\subsectionbegin#1{\vskip20pt\par\noindent{\bf#1}\par\vskip12pt}
\def\topic#1{\vskip5pt\par\noindent{\topicfont#1}\ \ \ \ \ }
     
\def\ltsim{\mathrel{\spose{\lower 3pt\hbox{$\mathchar"218$}}
     \raise 2.0pt\hbox{$\mathchar"13C$}}}
\def\gtsim{\mathrel{\spose{\lower 3pt\hbox{$\mathchar"218$}}
     \raise 2.0pt\hbox{$\mathchar"13E$}}}
     
\def\sec{\hbox{$^s$\hskip-3pt .}}
\def\gg{\hbox{$>$\hskip-4pt $>$}}
\parskip=3pt
\def\gapprox{$_ >\atop{^\sim}$}     
\def\lapprox{$_ <\atop{^\sim}$}     
\def\apequal{\mathrel{\spose{\lower 1pt\hbox{$\mathchar"218$}}
     \raise 2.0pt\hbox{$\mathchar"218$}}}
\newbox\grsign \setbox\grsign=\hbox{$>$} \newdimen\grdimen \grdimen=\ht\grsign
\newbox\simlessbox \newbox\simgreatbox
\setbox\simgreatbox=\hbox{\raise.5ex\hbox{$>$}\llap
     {\lower.5ex\hbox{$\sim$}}}\ht1=\grdimen\dp1=0pt
\setbox\simlessbox=\hbox{\raise.5ex\hbox{$<$}\llap
     {\lower.5ex\hbox{$\sim$}}}\ht2=\grdimen\dp2=0pt
\def\gtorder{\mathrel{\copy\simgreatbox}}
\def\ltorder{\mathrel{\copy\simlessbox}}
\def\simgreat{\mathrel{\copy\simgreatbox}}
\def\simless{\mathrel{\copy\simlessbox}}

\def\Kp{K$^\prime$\ }
\def\etal{{\it et al.\ }}
\title{\bf DISTANT GALAXY CLUSTERS IDENTIFIED
FROM OPTICAL BACKGROUND FLUCTUATIONS$^1$}
\author{Dennis Zaritsky\altaffilmark{2}, Amy E. Nelson\altaffilmark{2},
Julianne J. Dalcanton\altaffilmark{3}, 
 and Anthony H. Gonzalez\altaffilmark{2}}
\bigskip

\affil{$^2$ UCO/Lick Observatory \& Board of Astronomy and Astrophysics, }
\vskip -12pt
\centerline{Univ. of California at Santa Cruz, Santa Cruz, CA 95064;}
\vskip -10pt
\centerline{dennis@ucolick.org, anelson@ucolick.org,
anthonyg@ucolick.org}
\vskip 5pt
\affil{$^3$ Hubble Fellow. Observatories of the Carnegie Institution of Washington,
813 Santa }
\vskip -12pt
\centerline{Barbara St., Pasadena CA 91101; jd@ociw.edu}
\vskip 2in
\noindent
$^1$ Based on observations made at the 
W.M. Keck telescope which is operated jointly by the California
Institute of Technology and the University of
California, and the 60-inch
telescope at the Palomar Observatory, California Institute of
Technology.
\vskip 1in
\begin{abstract}
We present the first high redshift ($0.3 < z <
1.1$) galaxy clusters found by systematically identifying
optical low surface brightness fluctuations
in the background sky. Using
spectra obtained with the Keck telescope and I-band
images from the Palomar 1.5m telescope, we conclude that
at least eight of the ten candidates examined 
are high redshift galaxy clusters.
The identification of such clusters from low
surface brightness fluctuations provides a complementary alternative to 
classic selection methods based on overdensities of 
resolved galaxies, and enables us to search efficiently for rich
high redshift clusters over large areas of the sky.
The detections described here are the first 
in a survey that covers a total of nearly 140 sq. degrees of
the sky and should yield, if these preliminary results are
representative, over 300 such clusters. 
\end{abstract}

\section{Introduction}

The development of structure in the Universe is one of the 
principal unanswered questions in cosmology. Although the local
structure in the distribution of galaxies is now well delineated
(\eg the CfA survey; Davis \etal 1982, de Lapparent, Geller, \& Huchra 1986; 
and the LCRS, Shectman \etal 1996), 
the corresponding distribution of galaxies
at large redshifts (z $\gtsim$ 0.5)
is poorly constrained. Clusters are the most recognizable
signposts of structure, particularly at high redshifts, but 
there are relatively few known high redshift clusters.
The largest advance of the last decade in the
number of known distant clusters comes from the work of  
Postman \etal (1996), who presented a
large, well-defined cluster catalog. Their compilation contains 35 clusters
with estimated redshifts $\ge 0.5$, plus 25 clusters from an inhomogeneous
sample with estimated redshifts $\ge 0.5$ (the Postman \etal catalog 
includes clusters identified previously 
by Gunn, Hoessel, \& Oke 1986). Although this is the
largest available sample,
once one begins to divide the sample on the basis
of cluster properties, such as redshift or richness, there are few 
clusters per bin.
To confidently address issues of large-scale
structure, and to study cluster and galaxy evolution at high redshift 
in detail, 
larger, well-defined samples of distant clusters are essential. 

Compiling a significantly larger sample of high redshift clusters
using standard identification techniques is difficult because one needs
moderately deep, photometrically homogeneous images of a large area of the sky.
In previous optical surveys, clusters are
identified largely from statistical 
excesses of resolved individual galaxies 
on the sky. More sophisticated
approaches, such as that by Postman \etal \hskip -5pt , 
incorporate magnitude and
color filters to minimize the contamination from line-of-sight projections.
Deep two-color photometry requires the use of 
a large
telescope under photometric conditions and good seeing.
The large expenditure of time necessary to do such a project
on a 4 to 5-m class telescope 
makes it difficult to cover very large areas ($\gg 10$ sq. degree) of the sky.

An alternative approach to finding distant clusters
is described by Dalcanton (1996) and relies on detecting 
the light from the unresolved galaxies in a cluster.
The success of this
technique is predicated on the assumption that the total flux 
from these distant clusters is dominated by flux from 
unresolved cluster galaxies.
This is undoubtedly true in shallow exposures of high redshift clusters. 
The light from unresolved clustered galaxies combines to produce
a detectable surface brightness fluctuation in relatively shallow, but
intrinsically uniform, images of the sky.
Once one no longer needs to 
resolve many galaxies per cluster in order to identify clusters, 
one can (1) survey
a larger section of the sky 
because the requisite exposure time to detect a particular cluster is shortened
and/or (2) use small telescopes, for which it is possible
to obtain larger time allocations. Both of these
advantages enable one to efficiently survey a large area of the sky. 
Because the magnitude limit of our survey and that of
the Postman \etal are roughly similar ($\ltsim$ 1 mag difference), 
we optimistically expect that the
surface brightness detection technique will enable us
to identify a greater proportion of clusters at $z > 0.5$
than that found by Postman \etal  (their sample has an average
estimated redshift of 0.52 for clusters with estimated redshifts 
$\ge 0.3$). Finally, 
our method provides an independently selected
sample of distant clusters, which is critical for revealing selection
biases and determining completeness. The two approaches are complementary.

In this {\it Letter}
we discuss the selection of the first ten candidate clusters from low surface
brightness fluctuations in 
drift scan images and the
vetting of those candidates using spectroscopy plus
optical imaging.
Our results demonstrate that the technique based on surface brightness
fluctuations is a viable
and efficient way to detect clusters out to beyond $z = 1$.

\section{Observations and Data Reductions}

We select our initial set of cluster candidates from the
drift scan survey ($\sim$ 17.5 sq. degrees) described by Dalcanton
\etal 1997. Our candidates are
5$\sigma$ fluctuations over background and typically
have some visible signs of substructure, with which we crudely
differentiate them from low surface brightness galaxies. Our
reduction and analysis procedure
includes carefully flatfielding the scans, masking bright stars and
galaxies, filtering large
fluctuations
that are due to sky variations or Galactic cirrus, removing
individual
resolved stars and galaxies, smoothing the remainder with an exponential
kernel of size comparable to the expected core of distant
clusters ($\sim$ 5 to 10 arcsec), and identifying the statistically
significant
fluctuations that are not the result of a low surface brightness
galaxy, scattered light, or Galactic dust.
We have compiled a preliminary list of 52 such cluster 
candidates from these images. 

We are currently
reducing drift scans of another 120 square degrees of sky observed
at the Las Campanas Observatory, from which we will identify several
hundred such candidates, for our full survey. 
The results described here are being
used to 
develop more quantitative selection criteria and to train our cluster
detection algorithms. Because we are currently conservative in our
cluster candidate selection by requiring 
that the fluctuation also 
have visible signs of substructure, presumably caused by the few brightest
cluster galaxies, we are currently biased toward the lower portion of
the accessible redshift range. Therefore, the redshift
distribution of the clusters presented here
is a conservative estimate of the expected final 
redshift distribution. 
This sample is not
representative of the final catalog, but 
demonstrates the viability of the technique.
From our current list of candidates, we selected a set
of ten that appeared promising.  

We have obtained some combination of photometry and spectroscopy of the ten
candidate cluster fields and list those observations in Table 1.
The V and I-band images were obtained at the Palomar
1.5m telescope during May 9 - 14, 1996.
Typical exposures per object are between 0.5 and 1.5 hours and calibration
is done using Landolt (1983, 1992) standard fields. All the fields
were observed at least once during photometric conditions. The photometric
solutions for the standards show some slight scatter (0.06 mag) that
suggests that the nights were not entirely photometric. We propagate
this uncertainty, plus the internal uncertainty from 
the SExtractor (Bertin \& Arnouts 1996) photometry, through to the final
magnitudes. A listing of the 
magnitudes will be presented in our study of cluster galaxy
properties (Nelson \etal 1997).

Spectroscopic observations were done at the Keck telescope using
the LRIS spectrograph with a 600 line mm$^{-1}$ grating and a 
single long slit aperture on Dec 20-21, 1995. 
The aperture was oriented to include
as many individual galaxies as were visible on the guider image
(usually 2 to 3 to an estimated $m_r < 22$). 
The typical exposure time was 30 min. 
In the end, we obtain a measurable spectrum for 5 to 8 objects per
slit position.
The images are rectified and calibrated using calibration 
lamp exposures and the night sky lines (see Kelson \etal 1997 for details). 
Velocities are measured using a cross-correlation technique
or centroids of emission lines. The rms redshift difference among redshifts
measured from different emission lines in the same spectra is 
0.0033 (the typical difference is $< 0.0005$). In cases where only
one emission line is observed, we attribute it to H$\beta$. There is
no ambiguity between H$\beta$ and [O II] because the
[O II] line is resolved into the two components at 3726 and 3729\AA.
The distribution of galaxy velocities
in each of the 10 observed fields is shown in Figure 1. 
We define a clump of at least two galaxies within
1000\kms\ of the clump mean to be a candidate cluster.
In Table 1 we list the candidate cluster redshift using the ``richest''
redshift clump along the line of sight
and list the number of galaxies in that
clump, $N_G$. Admittedly, for 
cases where only two galaxies satisfy the 1000\kms\ criterion,
the identification of the redshift pair as indicative of a cluster 
is suspect (but see below for possible vindication). 

\section{Discussion}

Cluster classification can be enigmatic, even at low redshifts
(Zabludoff \etal 1993). At higher redshifts, with few redshifts per
putative cluster, the task of vetting candidates is even more 
difficult. To proceed,
we adopt 1000\kms\ as a canonical rich cluster velocity
dispersion (cf. Zabludoff \etal \hskip -5pt ).
Each of the line-of-sight redshift distributions
for the ten fields contains at least two galaxies within 1000\kms\ 
of each other at some redshift. Assuming a smooth parent
distribution of redshifts defined from the sum of all of our 
measured redshifts, there is a negligible formal chance of 
finding two galaxies within 1000\kms\ of each other from among the
five to seven observed galaxies in a single field. 
However, because galaxies are
spatially correlated, the probability of finding galaxies clumped
in redshift space must be greater that this calculation suggests
and must depend on the unknown correlation
function at high redshifts. 

We conservatively estimate the probability of identifying spurious
groups in redshift space directly from our data.
By assuming that the four candidate cluster fields
in which the richest redshift clump
contains only two galaxies are spurious, we infer that the probability
of finding a spurious pair in redshift along a random line of sight
is at least 0.4 (4 of 10). Adopting this value as
the probability of random pairs, we
then expect that at least one of these four lines of sight, and at 
least two of the other six lines of sight, will contain a second
random pair. However, only one
of the ten fields has a second clump of two galaxies,
suggesting that the random probability of pairs is actually less than 0.4 and 
that at least some of
the fields that have a two galaxy redshift clump are probably not
spurious
cluster detections.
A similar argument supports the conclusion 
that groupings of three
or more galaxies are exceedingly unlikely. Therefore, we consider clumps with
at least three members to be clusters.
We temporarily ignore candidates that consist of 
a clump of only two galaxies, but we note that at least some are {\it probably}
clusters. We conclude on the basis of the spectroscopy 
that {\it at least} six of 
the ten candidates are {\it bonafide} clusters.

Photometry can also be used examine cluster candidates.
For our nearest cluster candidates ($z < 0.5$), 
the presence of a cluster is 
evident from our moderately deep images. However, there are three quantitative
photometric signatures of a cluster that might be evident even for
the most distant clusters: (1) a spatial clustering of resolved galaxies,
(2) a sharp rise in the number counts of galaxies per magnitude at 
apparent magnitudes fainter than that of the brightest
cluster galaxy, and (3) a 
sharp rise in the number counts per $V-I$ color bin at colors 
redder than that of a passively evolving, old giant elliptical galaxy.

The radial distribution of galaxies relative to the center of the
original surface brightness fluctuation is
shown in the lower panel of Figure 2 for the eight candidates for
which we have $I$-band images.
In all cases, including the
candidates for which only two galaxies were found clumped along the
line of sight,
the galaxies cluster around the position of the original low
surface brightness feature. This confirms that we have identified 
low surface brightness features produced by a clustered statistical excess of
unresolved galaxies on the sky (as opposed to low surface
brightness galaxies or fluctuations in the sky emission).

The $I$-band luminosity functions for galaxies within 40 arcsec of the
candidate cluster centers are shown in the upper panel of 
Figure 2 for the same eight candidates. To crudely quantify the location of the
expected sharp rise in the number counts of galaxies if these are
indeed clusters,
we fit a straight line
to the left portion of each histogram. The fit spans
the region between the leftmost bin that has at least two galaxies and
the peak of the histogram. The magnitude at which the fit intercepts the
horizontal axis is referred to as  $m_I^0$. 
Monte-Carlo
simulations with random field centers illustrate that random fields
do not produce the signatures seen in Figure 2. 
First, random fields do not typically show any spatial
clustering (only $\sim$ 20\% of the fields showed any central concentration).
Second, random fields typically contain few ($< 10$) galaxies within 40 arcsec
and those
galaxies are often scattered in magnitude so that one is unable
to fit to the bright end of the distribution in the same manner as for
the cluster fields ($m_I^0$ could be measured in only $\sim$ 30\% of
the random fields, and the value ranged from 17.6 to over
20). Therefore,
both the spatial clustering and the number and luminosities of
galaxies in the target fields support the conclusion that the candidates are
clusters. The final piece of supporting evidence comes from the correlation
between $m_I^0$ and redshift shown in Figure 3. 

The relation between
$m_I^0$ and redshift involves a combination of cosmology and 
cluster/galaxy evolution. One approach is to calibrate
the relationship empirically. We
find a strong correlation between $m_I^0$ and redshift (Spearman correlation
coefficient of 0.940 and a 0.9995 probability that this correlation
is not a random
effect). Excluding the most distant candidate, which appears to lie
off the ``best''
linear relation (our images are insufficiently deep to resolve 
a significant number of cluster members at $z > 1$), 
we fit a line with an {\it rms}
redshift scatter of only 0.05. This result further confirms
that these eight candidate clusters are real and suggests that $m_I^0$
may be an excellent empirical cluster redshift indicator over this range of 
redshifts. A complementary way to determine the $m_I^0 - z$ 
relation is to assume no 
cluster or galaxy evolution (\ie assume that
m$_I^0$ is a standard candle) and simply apply K-corrections and
the redshift-distance relation
to $m_I^0$. This approach results in the dashed line in Figure 3 (for $q_0 =
0.2$, $\Lambda = 0$, 
and K-corrections drawn from Fukugita, Shimasaku, \& Ichikawa
(1995) for
elliptical galaxies). The plotted curve is normalized to the empirical line 
at $z = 0.3$. The agreement between the empirical and theoretical
curves further confirms that the majority of these systems are clusters.
Once we obtain a larger cluster sample, 
we will develop a more sophisticated version of this approach
(possibly
combined with color information) to photometrically measure redshifts 
for the majority of the candidates.

Lastly, the $V$-band data was of insufficient quality to provide unambiguous
information on the colors of the cluster galaxies in the more
distant cluster candidates, and so we do not discuss it here. The
galaxy colors for the nearer clusters are consistent with their
spectroscopic redshifts. 

\section{Summary}

We present the initial set of high redshift clusters found
by identifying clusters from low surface
brightness fluctuations in the background sky (Dalcanton 1996). By
analyzing follow-up spectroscopic and photometric observations of ten
candidates, we demonstrated that the technique is at least 
60\%, possibly $>$ 80\%, successful, 
even before we have an adequate training sample
in  hand to tune our definition of a candidate cluster. 
These, and future spectroscopically 
confirmed candidates, will form our classification training sample.
If the success rate found here is representative,
we expect to identify well over 300
clusters at $0.3 < z \ltsim 1.1$.

The advent of a large sample of clusters with $z > 0.5$ and possibly 
a significant sample at $z > 1$ opens up detailed cluster and 
galaxy evolution studies based on 
carefully constructed samples of clusters, 
searches for supernovae at z $> 1$, and
searches for X-ray clusters at lower sensitivity levels than those
relying solely on X-ray detections. All of these will eventually lead
to a better understanding of galaxy and clusters evolution, and 
cosmology.

\vskip 0.5in
\noindent
Acknowledgments: DZ and AEN gratefully acknowledge funding from the California Space
Institute. AEN also acknowledges support from a UCSC Graduate Research
Mentorship Fellowship. Financial support for JJD was provided by NASA through
Hubble Fellowship grant \#2-6649 awarded by the Space Telescope
Science Institute, which is operated by the AURA, Inc., for NASA
under contract NAS 5-26555. AHG acknowledges support from an 
NSF Graduate Research Fellowship.

\clearpage

\noindent
{\bf References}
\aa{Bertin, E., \& Arnouts, S. 1996}{117}{393}
\apj{Dalcanton, J.J. 1996}{466}{92}
\refbook{Dalcanton, J.J., Spergel, D.N., Gunn, J.E., Schmidt, M., 
\& Schneider, D.P. 1997, AJ, submitted}
\apj{Davis, M., Huchra, J., Latham, D.W., \& Tonry, J. 1982}{253}{423}
\apjlett{de Lapparent. V., Geller, M.J., \& Huchra, J.P. 1986}{302}{L1}
\pasp{Fukugita, M., Shimasaku, K., \& Ichikawa, T. 1995}{107}{945}
\apj{Gunn, J.E., Hoessel, J.G., \& Oke, J.B. 1986}{306}{30}
\apjsup{Huchra, J., Davis, M., Latham, D., \& Tonry, J. 1983}{53}{89}
\refbook{Kelson, D.D., van Dokkum, P., Franx, M., \& Illingworth,
G.D. 1997, in prep.}
\aj{Landolt, A.U. 1983}{88}{439}
\aj{Landolt, A.U. 1992}{104}{340}
\refbook{Nelson, A.E., Zaritsky, D., Gonzalez, A.H., \& Dalcanton,
J.J. 1997, in prep.}
\aj{Postman, M., Lubin, L. Gunn, J.E., Oke, J.B., Hoessel, J.G.,
Schneider, D.P., \& Christensen, J.A. 1996}{111}{615}
\apj{Shectman, S.A., Landy, S.D., Oemler, A., Tucjer, D., Lin, H.,
Kirshner, R.P., \& Schechter, P.L. 1996}{470}{172}
\aj{Zabludoff, A.I., Geller, M.J., Huchra, J.P., \& Ramella, M. 1993}{106}{1301}

\clearpage
\centerline{\bf Figure Captions}

\noindent
Figure 1 - The redshift for every detected galaxy in each of the 
ten fields is plotted. The order of the fields from top
to bottom matches that in Table 1. The filled circles denote galaxies
within the 1000\kms\ groupings. 

\bigskip
\noindent
Figure 2 - The I-band luminosity function (top) and radial surface
density profiles (bottom)
are plotted for the eight cluster candidates (Nos. 1, 2, 3, 4, 6, 7, 8, 9 from
left to right) for which we have $I$-band images. 
The lines drawn 
in the upper panel are least-squares fits to the left-hand portion
of the histograms from the first bin that has at least 2 galaxies to
the peak of the histogram.

\bigskip
\noindent
Figure 3 - The intercept from Figure 2, $m_I^0$, is plotted
against redshift. The triangles denote
candidates with only two galaxies identified in a redshift clump.
The solid line is a least-squares fit to the clusters at $z < 1$. The {\it
rms} about this line for $z < 1$ clusters is 0.05. The dotted line
represents the expected magnitude-redshift relationship for a $q_0 =
0.2$, $\Lambda = 0$
universe, where the curve has been normalized to coincide with
the empirical curve at $z = 0.3$.

\bigskip
\noindent

\begin{table}
\caption{Cluster Candidates}
\bigskip
\plotone{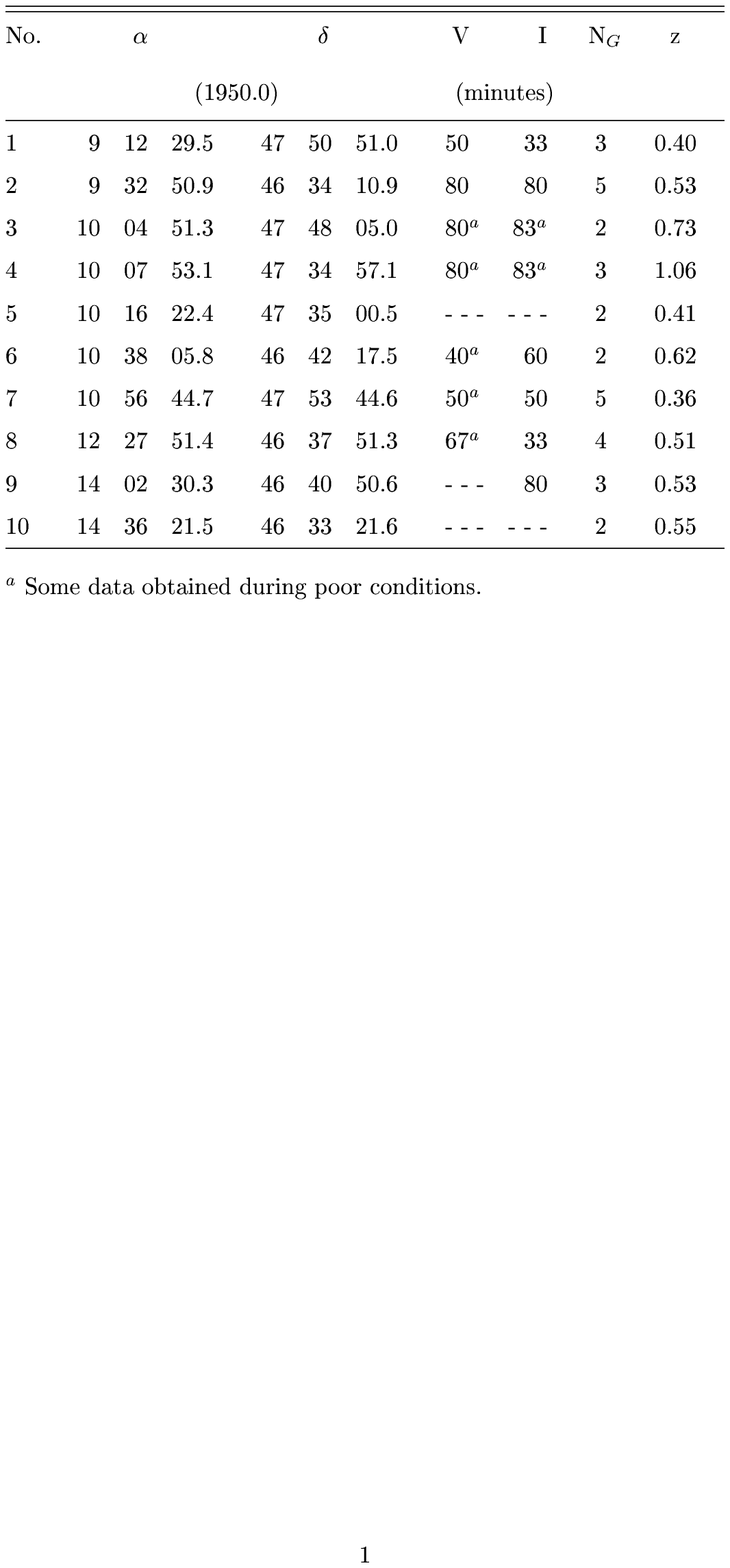}
\end{table}

\begin{figure}
\caption{}
\plotone{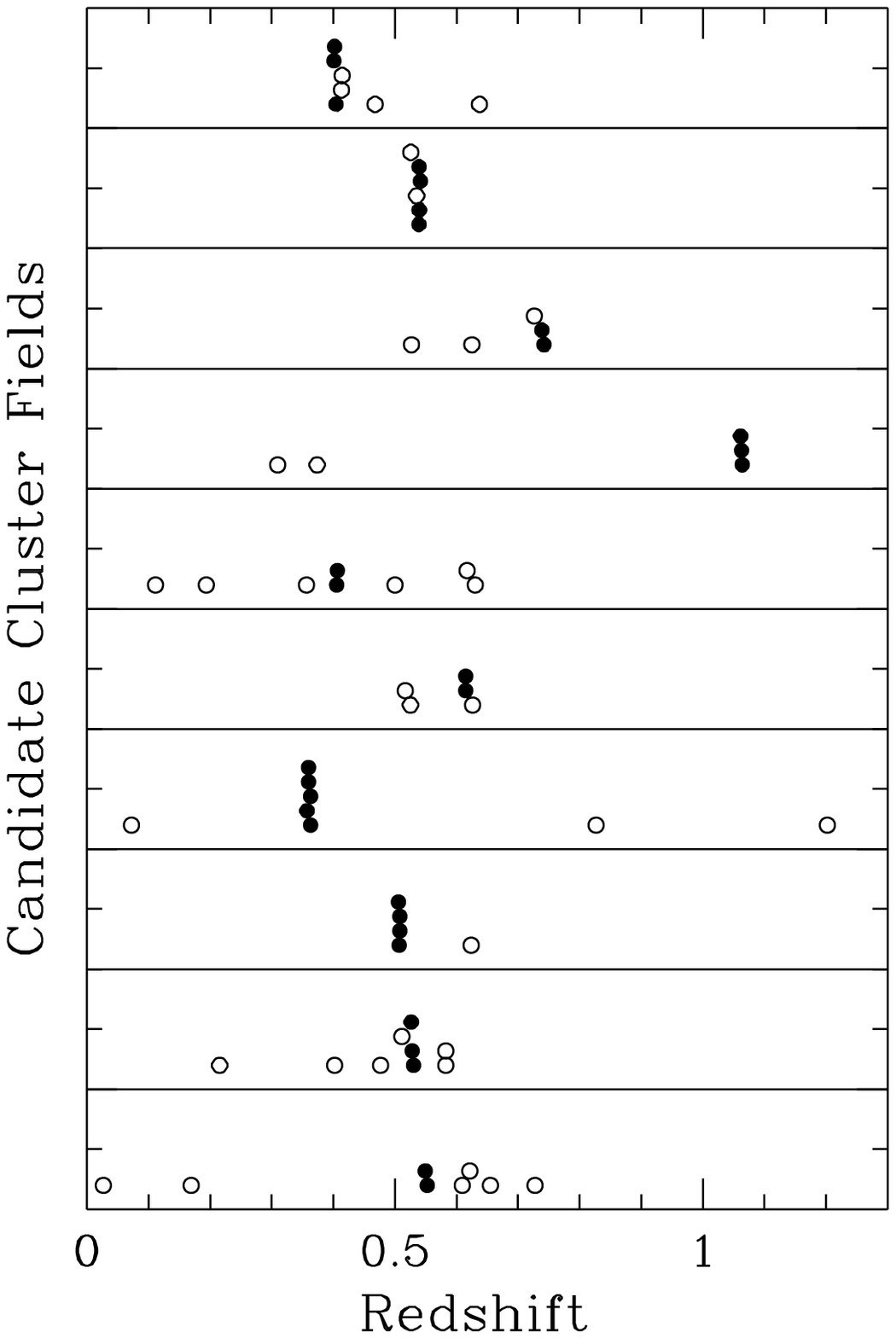}
\end{figure}
\clearpage

\begin{figure}
\caption{}
\plotone{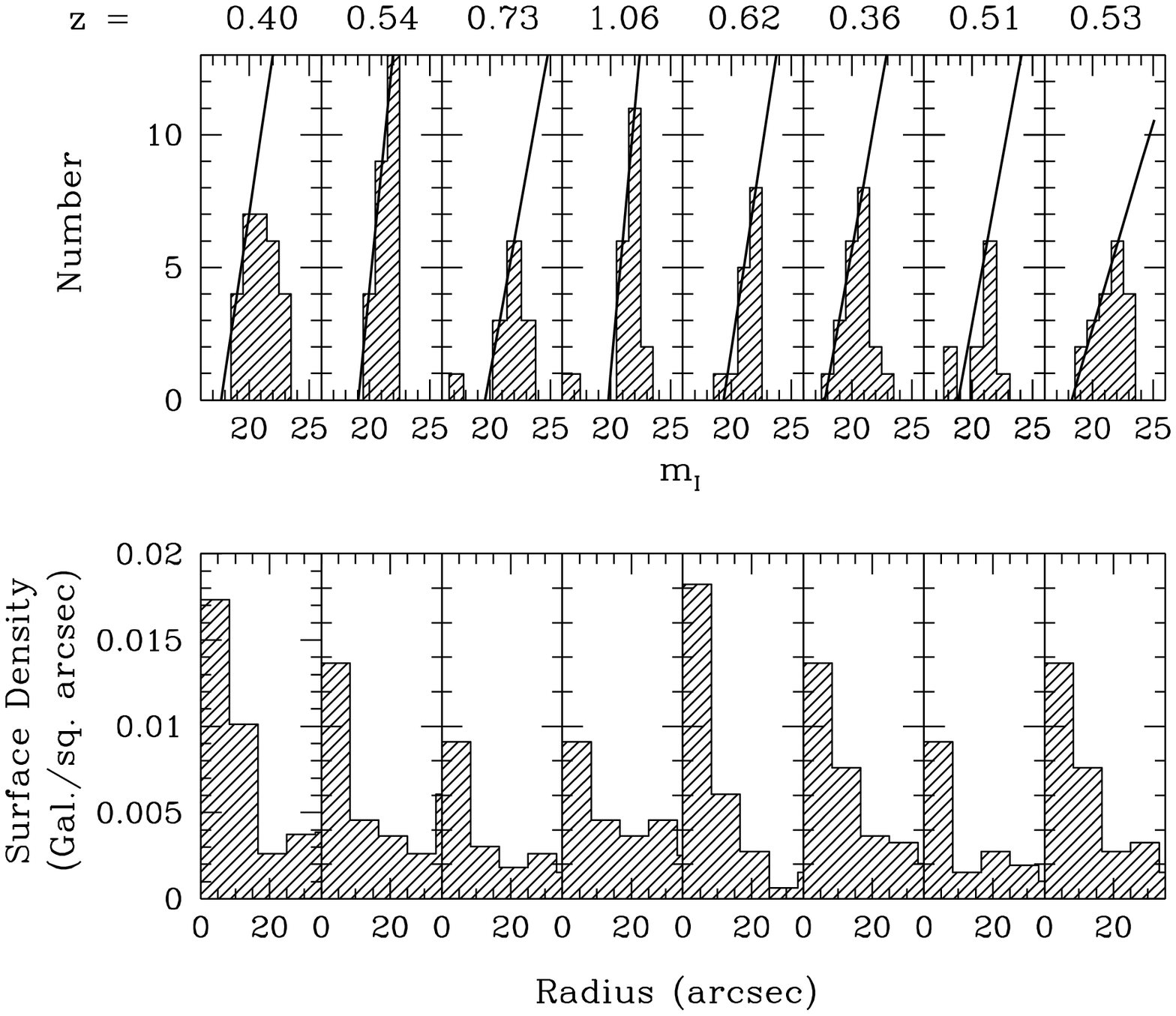}
\end{figure}
\clearpage

\begin{figure}
\caption{}
\plotone{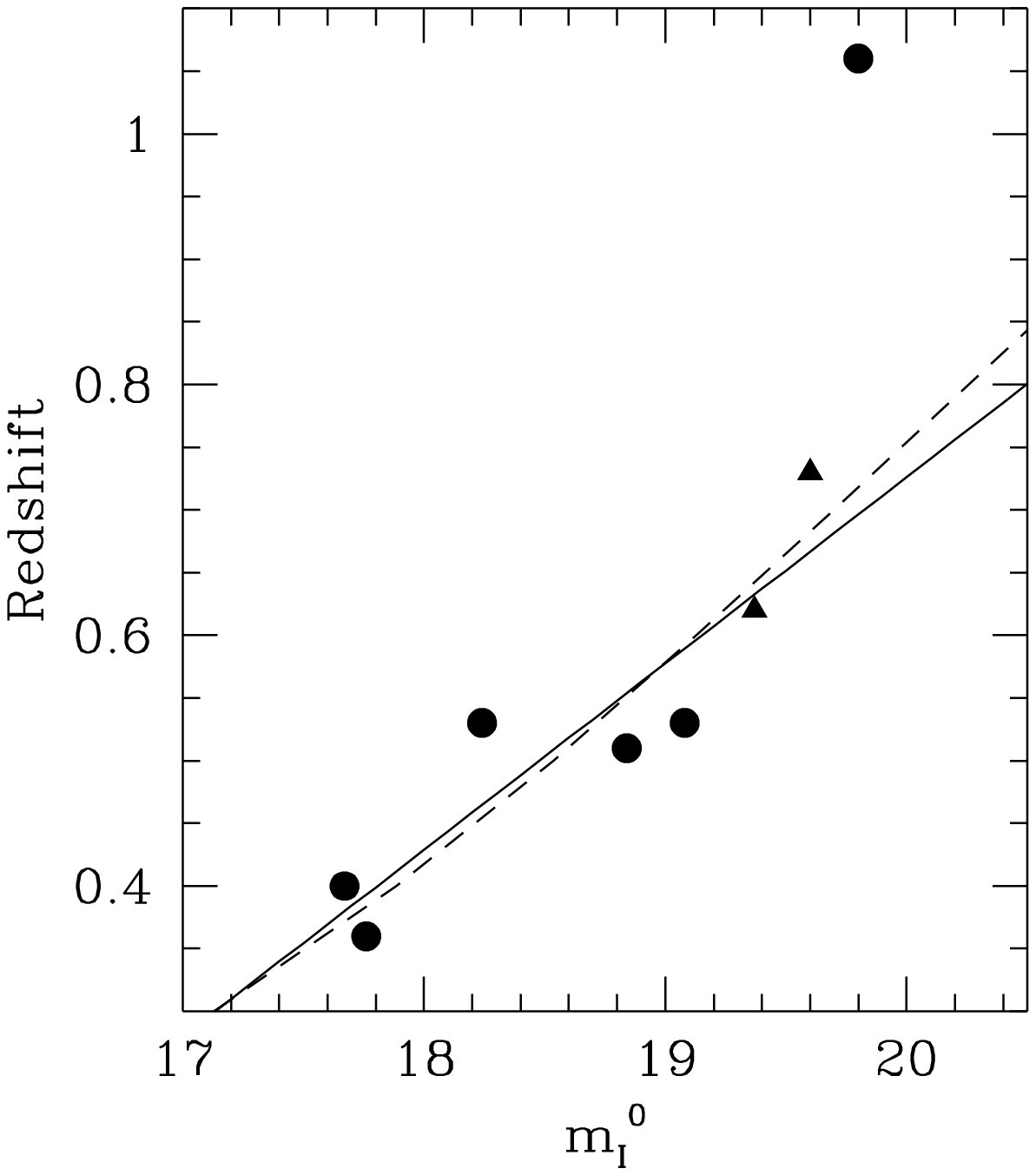}
\end{figure}
\clearpage

\end{document}